\documentclass[10pt]{iopart}

\expandafter\let\csname equation*\endcsname=\relax
\expandafter\let\csname endequation*\endcsname=\relax

\usepackage{iopams}
\usepackage{graphicx}
\usepackage{amsthm}
\usepackage{empheq}
\usepackage{color}
\usepackage{mathtools}
\usepackage{epstopdf}
\usepackage{cite}

\newtheorem{Theorem}{Theorem}[section]
\newtheorem{Proposition}[Theorem]{Proposition}

\newtheorem{Lemma}[Theorem]{Lemma}
\newtheorem{Corollary}[Theorem]{Corollary}

\theoremstyle{definition}

\newcommand*{\mrm}{\textrm}
\newcommand{\bTheorem}[1]{
\begin{Theorem} \label{T#1} }
\newcommand{\eT}{\end{Theorem}}

\newcommand{\bProposition}[1]{
\begin{Proposition} \label{P#1}}
\newcommand{\eP}{\end{Proposition}}

\newcommand{\bLemma}[1]{
\begin{Lemma} \label{L#1} }
\newcommand{\eL}{\end{Lemma}}
\newcommand{\bCorollary}[1]{
\begin{Corollary} \label{C#1} }
\newcommand{\eC}{\end{Corollary}}

\allowdisplaybreaks

\usepackage{xcolor}

\def\dt{\partial_t}
\def\na{\nabla}

\def\div{\operatorname{div}}

\def\u{\mathbf{u}}

\def\Dv{\mathbf{D}_S\mathbf{v}}

\def\la{\left\langle}
\def\ra{\right\rangle}
\def\dt{\partial_t}
\def\E{\mathcal{E}}

\newcommand{\Uc}[1]{\overset{\triangledown}{#1}}
\renewcommand{\u}{\mathbf{u}}
\def\vv{\mathbf{v}}
\def\ww{\mathbf{w}}

\def\CC{\mathbf{C}}
\def\DD{\mathbf{D}}
\newcommand{\I}{\mathbf{I}}
\newcommand{\TT}{\mathbf{T}}
\newcommand{\trr}[1]{\mathrm{tr}({#1})}
\newcommand{\trC}{\mathrm{tr}(\CC)}
\renewcommand{\div}[1]{\mathrm{div}\left({#1}\right)}

\newcommand{\dib}[1]{\mathrm{div}\big({#1}\big)}

\newcommand{\fd}{\frac{\mathrm{d}}{\mathrm{d}t}}
\newcommand{\dx}{\,\mathrm{d}x}

\DeclarePairedDelimiter{\norm}{\|}{\|}
\DeclarePairedDelimiter{\snorm}{|}{|}

\def\softd{{\leavevmode\setbox1=\hbox{d}%
          \hbox to 1.05\wd1{d\kern-0.4ex{\char039}\hss}}}

\begin{document}

\title{Analysis of a viscoelastic phase separation model}

\author{Aaron Brunk$^1$,
  Burkhard D\"unweg$^{2,3}$,
  Herbert Egger$^{4}$,
  Oliver Habrich$^4$,
  M\'aria Luk\' a\v{c}ov\'a-Medvi\softd ov\'a$^1$
  and Dominic Spiller$^2$}

\address{$^1$ Institute of Mathematics, Johannes Gutenberg University Mainz,
  Staudingerweg 9, 55128 Mainz, Germany \\
  $^2$ Max Planck Institute
  for Polymer Research, Ackermannweg 10, 55128 Mainz, Germany\\
  $^3$ Department of Chemical Engineering, Monash University, Clayton,
  Victoria 3800, Australia\\
  $^4$ Department of Mathematics, Technical
  University Darmstadt, Dolivostraße 15, 64293 Darmstadt, Germany}

\ead{abrunk@uni-mainz.de, duenweg@mpip-mainz.mpg.de,
  egger@mathematik.tu-darmstadt.de,
  habrich@mathematik.tu-darmstadt.de,
  lukacova@mathematik.uni-mainz.de, spiller@mpip-mainz.mpg.de}

\begin{abstract}
  A new model for viscoelastic phase separation is proposed, based on
  a systematically derived conservative two-fluid model. Dissipative
  effects are included by phenomenological viscoelastic terms. By
  construction, the model is consistent with the second law of
  thermodynamics. We study well-posedness of the model in two space
  dimensions, i.e., existence of weak solutions, a weak-strong
  uniqueness principle, and stability with respect to perturbations,
  which are proven by means of relative energy estimates. Our
  numerical simulations based on the new viscoelastic phase separation
  model are in good agreement with physical experiments. Furthermore,
  a good qualitative agreement with mesoscopic simulations is
  observed.
\end{abstract}

\noindent{\it Keywords}: viscoelastic phase separation,
relative energy, weak-strong uniqueness, dynamic structure factor

\submitto{\JPCM}

\ioptwocol

\section{Introduction}

In this work we propose and analyze a new viscoelastic phase
separation model.

The mixing properties of polymer solutions are highly
temperature-dependent, leading to the terms ``good'' and ``poor''
solvent. While good solvent corresponds to temperatures at which
contacts between solvent particles and macromolecules are more
favorable than self-contacts, for a poor solvent the opposite is the
case, i.e. self-contacts are more favorable. Of principle interest is
to understand the dynamics of demixing after a quench from good to
poor solvent. The relevant processes and dynamical equations are
well-known for Newtonian binary fluids, and typically described in
terms of the so-called ``model H''~\cite{Hohenberg.1977,
  onukiPhaseTransitionDynamics2002a,Cates2017}. However, in polymer
solutions several new effects appear which are related to the
different length and time scales of the solvent and the polymer
molecules, leading to so-called \emph{dynamical asymmetry}, see
\cite{Tanaka2017}.

The origin of our investigations goes back to the work of
Tanaka~\cite{Tanaka.}, who studied dynamic asymmetry theoretically and
experimentally, see also \cite{Araki2001,Zhang2001} for
three-dimensional simulations. The apparent inconsistency of Tanaka's
model with the second law of thermodynamics was addressed in
\cite{Zhou.2006} and cured by modifications based upon applying basic
principles of non-equilibrium
thermodynamics~\cite{Groot.2016,Grmela.1997,Grmela.1997b}. In this
work, a closure of the system was obtained by relating the relative
velocity between solvent and polymer through an ad-hoc constitutive
relation and additional equations for the viscoelastic stresses.  For
further work on viscoelastic phase separation we refer the reader to
Refs.~\cite{Doi, Helfand1989, Milner1993, Taniguchi1996, Onuki}, while
more recent studies including applications can be found in
\cite{Tsurusawa2017,Koyama2018,Surendran2020,PajicLijakovic2020,
  PajicLijakovic2021,Luo2017,Zhu2016}.

In Section \ref{sec:reduction}, we discuss an alternative derivation
starting from a conservative model for a binary fluid which can be
obtained by systematic coarse graining from a microscopic description
\cite{spiller2021systematic}.
Viscous effects are introduced via additional stresses of
phenomenological type, and some standard simplifications lead to a
simplified version of the model in \cite{Zhou.2006}. The systematic
derivation allows us to describe in detail the assumptions underlying
the model and to justify the closure relations used in
\cite{Zhou.2006}.
Motivated by the considerations in \cite{doi:10.1137/060666810}, we
include additional dissipative terms in the momentum equations and
utilize a change of viscoelastic variables to obtain our final model.
These deviations from \cite{Zhou.2006} turn out to be essential for
the verification of mathematical well-posedness.

In Section \ref{sec:macro} we discuss the consistency with the second
law of thermodynamics, and the existence of weak solutions. Based on
relative energy estimates, we study conditional uniqueness and
stability of solutions with respect to perturbations in the problem
data and the model parameters.

In Section~\ref{sec:comparison} we present a comparison of structure
factors obtained by simulations of the new model with those of a
mesoscopic description. A good qualitative agreement is observed in
these numerical tests, which may serve as a first validation. We also
refer to \cite{Brunk.08.07.2019,Brunk.30.04.2020} where results of
numerical simulations in two and three space dimensions are compared
with physical experiments from \cite{Tanaka.} and indicate very good
agreement.

\section{Model derivation}\label{sec:reduction}

We start from the following special case of a binary fluid model
proposed in \cite{spiller2021systematic}, namely
\begin{align}
\dot c &= -\div{\frac{1-c^2}{2 }\u} \nonumber\\
\dot \vv &= -\na p - \div{\frac{1-c^2}{4}\u\otimes\u}
\nonumber \\
0&=\div{\vv}\label{eq:binary} \\
\dot \u &= -2\na\frac{\partial f(c)}{\partial c}- (\u\cdot\na)\vv
+ c(\u\cdot\na)\u \nonumber \\
& + \frac{1}{2}(\u\otimes\u)\cdot\na c.
\nonumber
\end{align}
Here $c, \vv, \u, p$ denote, respectively, the mixture density
difference ($-1\leq c \leq 1$), the mass-averaged velocity, the
relative velocity, and the pressure, which here is the Lagrange
multiplier associated to incompressibility of $\vv$.
Furthermore, $f$ is the Helmholtz free energy density of mixing and
$\frac{\partial f}{\partial c}$ the corresponding chemical potential.
As usual, we denote by $\dot g = \partial_t g + (\vv\cdot\nabla)g$ the
convective derivative with respect to the mass-averaged velocity
$\vv$.

We assume in the following that $\Omega\subset \mathbb{R}^d$ is a
bounded domain and require, for simplicity of presentation,
appropriate periodic boundary conditions for all field variables. Then
\begin{equation}
  \fd \int_\Omega \left( E_{\text{kin}} + f(c) \right) dx = 0
\end{equation}
with $E_{\text{kin}} = \frac{1}{2}\snorm*{\vv}^2 +
\frac{1-c^2}{8}\snorm*{\u}^2$ denoting the kinetic energy
density. From a thermodynamic point of view, model \eqref{eq:binary}
is thus fully conservative, as it is appropriate for isothermal
conditions. Note also that we consider the overall density to be
unity.

In a phenomenological manner, we now add viscoelastic effects by
introducing corresponding stress tensors in the two momentum
equations. Thus the modified momentum equations read
\begin{align}
\dot \vv =& -\na p + \div{\TT_\vv}-
\div{\frac{1-c^2}{4}\u\otimes\u} \\
\dot \u =& -2\na\frac{\partial f(c)}{\partial c}
+ n(c)\div{\TT_u + q\I}- (\u\cdot\na)\vv\nonumber\\
& + c(\u\cdot\na)\u + \frac{1}{2}(\u\otimes\u)\cdot\na c .
\nonumber
\end{align}
Here $\TT_\vv$ and $\TT_\u$ denote the tensors related to the
trace-free parts of the (symmetrized) velocity gradients of $\vv$ and
$\u$, respectively, while $\TT_\u^{trace}=q\I$ is the tensor related
to the trace of the velocity gradient of $\u$. The tensor related to
the trace part of the velocity gradient of $\vv$ can be neglected,
since we assumed $\vv$ to be incompressible. Note that the momentum
balance requires a divergence form in the equation for $\vv$. The
divergence form of the viscoelastic variables in the equation for $\u$
is postulated.

Constitutive relations for the viscoelastic variables are required to
close the system.

As a next step, we introduce some simplifying assumptions: Since the
relative velocity $\u$ is a fast non-hydrodynamic variable, we assume
that $\u$ is quasi-static, and replace $\partial_t\u + (\vv\cdot\na)\u
+ (\u\cdot\na)\vv$ by a relaxation term
$\frac{1}{\tau}\u$. Furthermore, we assume $\u$ to be small and
neglect terms of higher order in $\u$. This leads to the simplified
momentum equations
\begin{align}
  \dot \vv &= -\na p + \div{\TT_\vv}\nonumber \\
  \frac{1}{\tau}\u
  &= -2\na \frac{\partial f(c)}{\partial c}
  + n(c)\div{\TT_u + q\I}.
  \label{eq:closure_rel}
\end{align}
As a consequence of the simplifying assumptions, the kinetic energy
will now only depend on $\vv$.

As a next step, we neglect the stress tensor $\TT_\u$ and, similar to
\cite{doi:10.1137/060666810,Zhou.2006}, we choose the dissipative
Oldroyd-B-type model to describe the time evolution of $q$ and
$\TT_\vv$.
In that case, the evolution of a stress tensor $\TT$ is described by
\begin{equation}
 \Uc{\TT} = -\frac{1}{\tau_\TT}\TT
+ \varepsilon\Delta\TT,	 \nonumber
\end{equation}
where $\Uc{\TT} = \partial_t\TT + (\vv\cdot\nabla)\TT - (\nabla\vv)\TT
- \TT(\nabla\vv)^\top$ is the upper convected derivative. For the
special case of a tensor $\TT=q\I$, one simply obtains, due to
incompressibility,
\begin{align}
\dot q = - \frac{1}{\tau_q}q  +\varepsilon\Delta q. \nonumber
\end{align}
By combination of the previous considerations and adding additional
contributions corresponding to the linear stresses $A \div{\u}$ and
$A_2 \Dv$ induced by the associated velocities, see \cite{Zhou.2006},
where $\Dv$ denotes the symmetrized velocity gradient tensor, we
arrive at the following intermediate model
\begin{align}
  \dot c &= -\div{\frac{1-c^2}{2}\u}, \quad
  \dot \vv = -\nabla p + \div{\TT_\vv}\nonumber, \\
  \frac{1}{\tau}\u &=
  -2\na\frac{\partial f(c)}{\partial c} + n(c)\div{q\I},
  \nonumber	\\
  \dot{q} &= - \frac{1}{\tau_q}q+ \varepsilon_1\Delta q + A\div{\u}, \\
  \Uc{\TT}_\vv &= -\frac{1}{\tau_\TT}\TT_\vv
  + \varepsilon_2\Delta\TT_\vv + A_2 \Dv .\nonumber
\end{align}
Following \cite{Zhou.2006} and common practice in the rheological
literature, we assume that $\TT_\vv$ is positive semidefinite,
i.e. has non-negative trace. Similarly, again following
\cite{Zhou.2006}, we postulate the free energy
\begin{equation}
  E = \int_\Omega \left[ \frac{1}{2}\snorm{\vv}^2 + f(c)
    + \frac{1}{2}q^2 + \frac{1}{2}\trr{\TT_\vv} \right] dx ,
  \label{eq:red_energy}
\end{equation}
whose time evolution is found, by straightforward calculation, to be
\begin{align}
  &\frac{\mathrm{d}}{\mathrm{d}t} E =
  - \frac{1}{\tau_q}q^2 - \varepsilon_1|\nabla q|^2 -
\frac{1}{\tau_\TT}\trr{\TT_v} \nonumber\\ -
&\frac{\tau}{n(c)}\left(\frac{1-c^2}{2}
\nabla\frac{\partial f}{\partial c}-
\nabla(Aq)\right)\left(\frac{2}{n(c)}
\nabla\frac{\partial f}{\partial c}- \nabla q\right) \nonumber.
\end{align}
By setting $A=1$ and $n(c)=4(1-c^2)^{-1}$, the last term becomes a
square, such that we obtain
\begin{align}
  \frac{\mathrm{d}}{\mathrm{d}t} E
  = &- \frac{1}{\tau_q}q^2 - \varepsilon_1|\nabla q|^2
  - \frac{1}{\tau_\TT}\trr{\TT_v} \nonumber\\
  &- n(c)\tau\snorm*{\frac{1-c^2}{2}
    \nabla\frac{\partial f}{\partial c}- \nabla q}^2 \leq 0,
  \nonumber
\end{align}
which shows that, under the above assumptions, the resulting model
is consistent with the second law of thermodynamics.
After eliminating the relative velocity $\u$, we arrive at the
following system
\begin{align}
  \dot c &= \tau\div{(1-c^2)\na\frac{\partial f(c)}{\partial c}- 2\na q}
  \nonumber\\
  \dot \vv &= -\nabla p + \div{\TT_\vv}, \; \div{\vv}=0 \nonumber \\
  \dot{q} &= - \frac{1}{\tau_q}q+ \varepsilon_1\Delta q +
  \tau\div{\frac{4}{1-c^2}\na q
    - 2\na\frac{\partial f(c)}{\partial c}} \nonumber\\
  \Uc{\TT}_\vv &= -\frac{1}{\tau_\TT}\TT_\vv
  + \varepsilon_2\Delta\TT_\vv + A_2 \Dv .
  \label{eq:red_full}
\end{align}
Let us note that the first and third equation form a cross-diffusion
system for $c$ and $q$, and that such systems are well-suited to
describe pattern formation.

We further observe that the structure of the reduced model
\eqref{eq:red_full} is exactly the same as that in
\cite{Zhou.2006}. In fact, the model of this reference can be obtained
by the following minor modifications:
\begin{enumerate}
\item change of variable $c$ to volume fraction $\phi$;
\item Newtonian part in the viscous stress in $\eqref{eq:red_full}_2$;
\item inclusion of the Korteweg interface stress for $\phi$ in
  $\eqref{eq:red_full}_2$;
\item extra stress $q$ replaced by $A(\phi)q$;
\item special choice of $\tau$ dependent on $\phi$;
\item $\varepsilon_i=0$ for $i=1,2$.
\end{enumerate}
The first two points are straightforward and we omit a thorough
discussion.
Point three can be incorporated by extending the potential $f$ to
include a gradient term modelling the interface stress, leading to the
typical Ginzburg-Landau form.
Point four is based on the idea that the bulk stress $q\I$ should
change rapidly when $\phi$ transits between the pure
concentrations. In fact, \cite{Tanaka.} assumed that for the solvent
$A$ should vanish. In fact, the whole derivation above can be done by
replacing $q$ by $A(\phi)q$ in the relation of the relative velocity,
i.e. \eqref{eq:closure_rel}.
Point five can be motivated on a similar reasoning.
Let us finally comment on the last modification: In most of the
rheological literature, non-diffusive equations with
$\varepsilon_i=0$ are considered. A justification for
$\varepsilon_i>0$ has been given in \cite{doi:10.1137/060666810},
where it is argued that such terms are related to the center-of-mass
diffusion of the polymer chains. While the parameters $\varepsilon_i$
may be rather small in general, their positive value leads to much
stronger results concerning analysis and numerical treatment.

Including the modifications (i)--(v) will lead to the model of
\cite{Zhou.2006}, and our derivations so far may be considered as an
alternative derivation of their model.
On the basis of the same arguments, we will further replace the
Oldroyd-B model for the elastic stress tensor $\TT_\vv$ by the
Peterlin model for the conformation tensor $\CC$, which can be
interpreted as a nonlinear, diffusive Oldroyd-B model, where we assume
that the stiffness of the spring is not constant but depends on the
elongation of the conformation tensor, i.e., on $\trC$; see
\cite{LukacovaMedvidova.2017c} for details and the references therein.
Let us mention that, in analogy to the binary fluid model, the
dissipative viscoelastic phase separation model can be derived in the
context of the GENERIC formalism; see \cite{Brunk.08.07.2019} for
details. Hence our final model for viscoelastic phase separation reads
\begin{align}
  \dot \phi &= \dib{n^2(\phi)\nabla\mu -
    n(\phi)\nabla\big(A(\phi)q\big)}
  \label{eq:full_model}\\
  \dot q  &= -h_1(\phi)q + A(\phi)\div{\nabla\big(A(\phi)q\big)
    - {n(\phi)\nabla\mu}}
  \nonumber \\
  & + \varepsilon_1\Delta q
  \nonumber\\
  \dot \vv  &= \div{\eta(\phi)\Dv}
  - \na p + \dib{\TT}+ \mu\na\phi
  \nonumber\\
  \Uc{\CC}  &=  - h_2(\phi)B(\trC)(\trC\CC-\I)
  + \varepsilon_2\Delta\CC
  \nonumber	\\
  \mu &= -c_0\Delta\phi + \frac{\partial f(\phi)}{\partial \phi},
  \quad \div{\vv}=0,
  \quad \TT=\trC\CC.
  \nonumber
\end{align}
Here $\phi$ denotes the volume fraction of the polymers while the
other state variables have the same meaning as above. Furthermore,
$n(\phi)$ denotes the mobility function and $h_1$ is the relaxation
rate for $q$. Similarly, $h_2(\phi)B(\trC)$ denotes a
generalized relaxation rate, $\eta(\phi)$ is a volume fraction
dependent viscosity, and $A(\phi)$ is related to the dynamic
asymmetry, sometimes called bulk relaxation modulus; see
\cite{Tanaka.,Zhou.2006}.
Let us further mention that the elastic stress tensor $\TT$ is now
defined implicitly via an evolutionary equation for the conformation
tensor $\CC$, which is assumed to be positive-definite.

The term $\mu\nabla\phi$ is related to the Korteweg interface stress
and can be obtained in this form by a suitable redefinition of the
pressure $p$.

\section{Mathematical properties}\label{sec:macro}

In the following, we summarize the most important mathematical
properties of our new model \eqref{eq:full_model} for viscoelastic
phase separation.  For ease of notation, we again assume that the
equations are complemented by appropriate periodic boundary
conditions.

\subsection{Thermodynamic consistency}

Let us start by investigating the thermodynamic consistency of the
proposed model.
The free energy of the system \eqref{eq:full_model} is given by
\begin{align}
  &E(\phi,q,\vv,\CC)
  =  \int_\Omega \left( \frac{c_0}{2}\snorm*{\na\phi}^2
  + f(\phi) \right) \dx + \int_\Omega \frac{q^2}{2} \dx
  \nonumber \\
  &+\int_\Omega \frac{1}{2}\snorm*{\vv}^2 \dx +
  \int_\Omega \left( \frac{1}{4}\trC^2 - \frac{1}{2}\trr{\ln(\CC)} \right) \dx ,
  \label{eq:free_energy}
\end{align}
which indicates the decomposition of the free energy via $E(\phi,q,\vv,\CC)=
E_{mix}(\phi) + E_{bulk}(q) + E_{kin}(\vv) + E_{el}(\CC)$ into several
distinct contributions.
By a lengthy but straightforward calculation \cite{Brunk.08.07.2019},
one finds the following energy-dissipation identity
\begin{align*}
  &\frac{\mathrm{d}}{\mathrm{d}t} E =
  -\int_\Omega \Big( \snorm*{n(\phi)\nabla\mu
    - \nabla(A(\phi)q)}^2 + h_1(\phi)q^2\\ & +
  \varepsilon_1|\nabla q|^2
  + \snorm{\eta^{1/2}(\phi)\Dv}^2
  + \frac{\varepsilon_2}{2}\snorm{\CC^{-1/2}\nabla\CC\CC^{-1/2}}^2\\
  &+ h_2(\phi)B(\trC)\trC\trr{\TT + \TT^{-1} - 2\I}
  \\
  & + \frac{\varepsilon_2}{2}\snorm{\nabla\trC}^2 \Big) \, \dx
  \leq 0,
\end{align*}
which establishes the consistency with the second law of thermodynamics.

\subsection{Existence of solutions}

In the following we review results of
\cite{Brunk.08.07.2019,Brunk.30.04.2020}, where we rigorously proved
the existence of global weak solutions for \eqref{eq:full_model} in
two space dimension. The proofs in these works are based on energy
functionals similar to \eqref{eq:free_energy}.
We will distinguish the following two particular cases: The first,
so-called regular case, assumes a smooth and strictly positive
mobility function $n(s)$ paired with a general polynomial-like double
well potential, including the usual Ginzburg-Landau potentials.
The second case is concerned with degenerate mobility functions, which
are allowed to vanish at the points of single phases, paired with
logarithmic potentials, like the physically relevant Flory-Huggins
potential.

\begin{Lemma}[Regular case, \cite{Brunk.08.07.2019}] \label{lem:reg}
  Under suitable assumptions on the parameter functions, there exists
  at least one global weak solution of \eqref{eq:full_model} for the
  Ginzburg-Landau type potential.
\end{Lemma}

\begin{Lemma}[Degenerate case, \cite{Brunk.30.04.2020}] \label{lem:deg}
  Under suitable assumptions on the parameter functions, there exists
  at least one global weak solution of \eqref{eq:full_model} for the
  Flory-Huggins type potential. Furthermore, the solution satisfies
  the physical bounds $\phi\in[0,1]$.
\end{Lemma}

Recall that weak solutions are, in particular, solutions in the sense
of distributions. Moreover, the spatial and temporal regularity is
sufficient such that the energy-dissipation structure
\eqref{eq:free_energy} is preserved; we refer to \cite{Brunk.NEWNEW}
for details.
Let us finally remark that similar existence result are not proven and
actually not expected to hold for the model in \cite{Tanaka.}.

\subsection{Relative energy estimates,
  weak-strong uniqueness, and stability of solutions}

In addition to pure existence of weak solutions, we now report on
conditional uniqueness and stability results which encode the
mathematical well-posedness of the model \eqref{eq:full_model}.
The basic tool for the investigations of this section are relative
energy estimates.

We start by defining a modified free energy functional for the problem
under consideration, namely
\begin{equation}\label{eq:energy_fun}
  E =  E_{mix} + E_{bulk} + E_{kin} +
  \int_\Omega \left( \frac{1}{4}\snorm{\CC}^2
  + \frac{\alpha}{2} |\phi|^2 \right) \dx. \nonumber
\end{equation}
These modifications for the compositional and elastic degrees of
freedom are introduced in order to obtain a functional that is
strictly convex. This is the case if $\alpha > 0$ is sufficiently
large.

Let us recall \cite{Dafermos.1979,Dafermos.1979b} that for any
strictly convex functional $E(z)$, we may define a \emph{relative
  energy}
\begin{align}
  \mathcal{E}(z\vert\hat z) = E(z) - E(\hat z)
  - \left\langle E'(\hat z),z-\hat z\right\rangle.
  \label{eq:rel_energy_def}
\end{align}
This amounts to the quadratic Taylor remainder which, by strict
convexity of $E$, is proportional to the quadratic distance between
$z$ and $\hat z$. Note that the brackets denote a suitable inner
product or dual pairing.
From the physical point of view this can also be interpreted as
statistical distance on the phase space with respect to the energy
landscape or, more generally, as some measure for information.

Abbreviating the states by $z=(\phi,q,\vv,\CC)$ and $\hat z = (\hat
\phi,\hat q,\hat\vv,\hat\CC)$, the relative energy of our phase
separation model reads
\begin{align}
  \mathcal{E}(z|\hat z) & =
  \int_\Omega \Big(\frac{c_0}{2}\snorm{\na\phi-\na\hat\phi}^2
   + f(\phi|\hat \phi) +
  \frac{1}{2}\snorm*{q-\hat q}^2
  \nonumber\\
   &+\frac{1}{2}\snorm*{\vv-\hat\vv}^2 +
  \frac{1}{4}\snorm{\CC-\hat\CC}^2
  + \frac{\alpha}{2}\snorm{\phi-\hat \phi}^2 \Big) \dx, \nonumber
\end{align}
where we used $f(\phi|\hat\phi)=f(\phi) - f(\hat \phi) - f'(\tilde
\phi)(\phi-\hat \phi)$ to abbreviate the Taylor remainder.
For $\alpha$ sufficiently large, the energy $E(z)$ can be seen to be
strictly convex, and the relative energy $\E(z|\hat z)$ satisfies
\begin{align}
\mathcal{E}(z\vert \hat z)\geq 0
\quad \text{and} \quad \mathcal{E}(z\vert \hat
  z)=0 \Longleftrightarrow z=\hat z. \label{eq:rel_eng_pos}
\end{align}
Under additional assumptions, one can show that the relative energy
allows to bound the squared distance, i.e., there exists a positive
constant $\lambda$ such that
\begin{equation*}
    \mathcal{E}(z|\hat z)
    \geq \frac{\lambda}{2}\|z - \hat z\|^2.
\end{equation*}

Using these properties, we can show the following decay estimated for
the relative energy in a similar fashion as in \cite{Brunk.NEWNEW}.
\begin{Theorem}\label{thm:rel_energy}
  Let $z=(\phi,q,\vv,\CC)$ be a weak solution of \eqref{eq:full_model}
  and $\hat z = (\hat \phi,\hat q,\hat\vv,\hat\CC)$ be another
  sufficiently smooth solution of \eqref{eq:full_model}. Then
  \begin{equation}
    \mathcal{E}(z|\hat z)(t) + \int_0^t\mathcal{D}(s) ds
    \leq  C\mathcal{E}(z|\hat z)(0), \label{eq:rel_en_gron}
  \end{equation}
  for some constant $C>0$ and with relative dissipation functional
  defined by
  \begin{align*}
    \mathcal{D}=\tfrac{1}{2}
    &\int_\Omega \Big( \snorm{n(\phi)\nabla(\mu-\hat\mu)
      - \nabla(A(\phi)(q-\hat q))}^2 \\
    &+ h_1(\phi)\snorm{q-\hat q}^2 +
    \varepsilon_1\snorm{\nabla(q-\hat q)}^2  \\
    &+ \eta(\phi)\snorm{\Dv - \mathrm{\mathbf{D}}_S\hat\vv}^2
    + \frac{\varepsilon_2}{2}\snorm{\na(\CC-\hat\CC)}^2\\
    &+ h_2(\phi)B(\trC) \trC\snorm{\CC-\hat\CC}^2  \Big) \dx.
  \end{align*}
\end{Theorem}
As a direct consequence of Theorem~\ref{thm:rel_energy}, we obtain the
following weak-strong uniqueness principle.
\begin{Corollary}\label{lem:wsu}
  Let $\hat z $ be a smooth solution of \eqref{eq:full_model}. Then
  any other weak solution $z$ of \eqref{eq:full_model} with the same
  initial values $z(0)=\hat z(0)$ coincides with $\hat z$ on the
  lifespan of the latter.
\end{Corollary}
This result can be obtained applying Theorem 3.3 since
$\mathcal{E}(z|\hat z)(0)=0$ by construction, see
\eqref{eq:rel_eng_pos}. This implies $E(z|\hat z)(t) = 0$ for
$t\in(0,T)$.
Weak solutions, provided by Lemma~\ref{lem:reg}, are thus unique, if
at least one of them is sufficiently regular.

Theorem~\ref{thm:rel_energy} further allows to immediately deduce
stability of solutions with respect to perturbations in the initial
data. The following slight modification of the argument allows to
study stability also with respect to further perturbations.

Before we state our results, let us note that \eqref{eq:full_model}
involved an additional variable $\mu = -c_0 \Delta \phi +
\frac{\partial f(c)}{\partial c}$, i.e., the chemical potential. Now
let $(z,\mu)$, with $z=(\phi,q,\vv,\CC)$ as before, denote a weak
solution of \eqref{eq:full_model}, and let $(\hat z, \hat\mu)$ be a
corresponding given set of sufficiently smooth functions. By inserting
$(\hat z,\hat \mu)$ into \eqref{eq:full_model}, we can define
residuals $r_i$, $i\in\{\phi,\mu,q,\vv,\CC\}$, according to
\begin{align}
  \la \dt\hat\phi,\psi \ra
  &+ \la \hat\vv\cdot\na\hat\phi,\psi \ra  \label{eq:pert_sys}\\
  &+\la n^2(\phi) \nabla\hat\mu -
  n(\phi)\na(A(\phi)\hat q),\na\psi \ra
  \nonumber \\
  &=: \la r_\phi, \psi \ra, \nonumber\\
  \la \hat\mu,\xi \ra &- c_0\la \na\hat\phi,\na\xi \ra
  - \la f'(\hat\phi),\xi \ra =: \la r_\mu, \xi \ra,\nonumber \\
  \la \dt\hat q,\zeta \ra &+ \la \hat\vv\cdot\na\hat q,\zeta \ra
  + \varepsilon_1\la \na\hat q,\na \zeta \ra
  + \la h_1(\phi)\hat q,\zeta \ra \nonumber\\
  &+ \la \na(A(\phi)\hat q) - n(\phi)\na\hat\mu, \na(A(\phi)\zeta)\ra
  \nonumber\\
  &=: \la r_q, \zeta \ra,\nonumber \\
  \la \dt\hat\vv,\ww \ra &+ \la (\hat\vv\cdot\na)\hat\vv,\ww \ra
  + \la \eta(\phi)
  \mathrm{\mathbf{D}}_S\hat\vv,
  \mathrm{\mathbf{D}}_S\ww \ra  \nonumber \\
  &+ \la \trr{\hat\CC}\hat\CC,\na\ww\ra - \la \na\hat\phi\hat\mu,\ww \ra
  =: \la r_\u, \ww\ra, \nonumber \\
  \la \dt\hat\CC,\DD \ra &+ \la (\hat\vv\cdot\na)\hat\CC,\DD \ra
  - \la (\na\hat\vv)\hat\CC + \hat\CC (\na\hat\vv)^\top, \DD\ra \nonumber\\
  &+ \varepsilon_2\la \na\hat\CC,\na\DD \ra
  \nonumber\\
  &+ \la h_2( \phi)B(\trC)\trC\hat\CC,\DD \ra \nonumber\\
  &- \la h_2(\phi)B(\trr{\hat\CC}),\trr{\DD} \ra
  =: \la r_\CC, \DD \ra. \nonumber
\end{align}
The brackets here denote the inner product in the space of
square-integrable functions $L^2(\Omega)$ or corresponding duality
products, and the variational characterizations of the residuals are
assumed to hold for all sufficiently smooth test functions
$(\psi,\xi,\zeta,\ww,\DD)$.
We now obtain the following generalization of Theorem
\ref{thm:rel_energy}.
\begin{Theorem} \label{thm:rel_energy_generalized}
  Let $(z,\mu)$ be a weak solution of \eqref{eq:full_model} and $(\hat
  z,\hat\mu)$ be given smooth functions with corresponding residuals
  $r_i$ as defined in \eqref{eq:pert_sys}. Then
  \begin{align}
    & \mathcal{E}(z|\hat z)(t) + \int_0^t\mathcal{D}(s) ds \nonumber \\
    &\leq  C\mathcal{E}(z|\hat z)(0)
     + C\int_0^t \Big( \|r_\mu\|_1^2
     + \sum_{i \neq \mu} \|r_i\|_{-1}^2 \Big) ds, \nonumber
  \end{align}
with $\norm{\cdot}_1$ and $\norm{\cdot}_{-1}$ denoting appropriate norms.
\end{Theorem}
In contrast to Theorem \ref{thm:rel_energy}, which allowed us to
``only'' study the effect of varying the initial conditions, Theorem
\ref{thm:rel_energy_generalized} permits us to investigate the full
spectrum of stability, in particular with respect to varying
parameters, but also with respect to aspects like model variability
and asymptotic limits. In the context of numerical analysis, a
discrete version of the above relative energy estimate allows to
derive convergence error estimates for appropriate numerical schemes.
Closely related to the above results are \emph{relative entropy
  estimates} in the context of coarse-graining, that are used to
measure the information loss when traversing through a hierarchy of
models. Appropriate versions of the above relative energy estimates
can also be applied for such an analysis by choosing the functions
$(\hat z,\hat \mu)$ appropriately.

\section{Comparison with a mesoscopic model}
\label{sec:comparison}

In this section we will compare results from our macroscopic model
with those of a mesoscopic, particle based model.

In the mesoscopic simulation approach, implemented in the simulation
package Espresso++~\cite{guzman_espresso_2019}, the polymer component
is modeled via a Kremer-Grest type~\cite{grest_molecular_1986}
bead-spring model. The quality of the solvent can be varied by
adjusting the attraction strength of the pair interaction based upon a
modified Lennard-Jones potential~\cite{soddemann_generic_2001}. This
potential defines the scales for length $\sigma_\mrm{LJ}$, energy
$\varepsilon_\mrm{LJ}$, and thereby time $t_\mrm{LJ} =
\sqrt{(m\sigma_\mrm{LJ}^2) / \varepsilon_\mrm{LJ}}$, where $m$ is the
particle mass. The polymers are dissipatively coupled to a
Lattice-Boltzmann (LB) solvent background in order to include
hydrodynamic interactions~\cite{dunwegPub068LatticeBoltzmann2009,
  tretyakov_improved_2017}.

The macroscopic simulations are based on the viscoelastic phase
separation model \eqref{eq:full_model} that is approximated by the
finite element method proposed in \cite{Brunk.08.07.2019}. We study
the problem on the periodic square $\Omega=[0,128]^2$. For this we
choose the following parameter set
\begin{align*}
  f(\phi)&= \phi\ln(\phi) + (1-\phi)\ln(1-\phi)
  + \frac{28}{11}\phi(1-\phi), \\
  n^2(\phi)&=\phi^2(1-\phi^2),
  h_1(\phi)=(50\phi^2)^{-1}, c_0=1, \\
  A(\phi)&=0.5(1 + \tanh\left( 10^3(\cot(\pi\phi^*)
  - \cot(\pi\phi)) \right),\\
  \eta(\phi) &= 2 + \phi^2, h_2(\phi)=(10\phi^2)^{-1},
  \varepsilon_1=0,\\
  \varepsilon_2 &= 10^{-2}, B(\trC)=\trC.
\end{align*}

Note that since we consider $\phi$ as the volume fraction the relevant
range is $\phi\in[0,1]$ and we chose $\phi^*$ as the mean value of the
initial data for $\phi$. Let us discuss the dimensionality. Since
$\phi$ is dimensionless, $[c_0]=L^2$. Furthermore, $[\eta],[n^2] =
L^2T^{-1}$ and $[h_1]=T^{-1}$. This yields the dimensionality
\begin{equation}
  [\eta][h_1^{-1}]=L^2 = [n^2][h_1^{-1}] = [c_0].
\end{equation}
Note that the parametric functions are chosen such that $\eta
h_1^{-1}, n^2h_1^{-1},c_0\sim 1$, meaning that all typical length
scales match, and are of the order of one lattice spacing.
Similarly we obtain matching timescales via
\begin{equation}
 h_1^{-1} \sim c_0\eta^{-1} \sim c_0 n^{-2} \sim 1 ;
\end{equation}
this motivates us to define the time unit via $t_{FE} = c_0\eta^{-1}$.
For typical simulation results we refer to \cite{Brunk.08.07.2019}.
We can clearly observe that the numerical simulations agree well with
the real experiment presented in \cite{Tanaka.}, see Figure~8 of
Ref.~\cite{Brunk.08.07.2019}. The experiment represents the separation
of a polystyrene-poly vinyl methyl ether (PS-PVME) mixture right after
the quench. It is inspected via video phase-contrast microscopy. For a
more detailed discussion of the PS-PVME experiments we refer to
\cite{Polios1997}.  According to this observation Tanaka introduced
six different regimes of the viscoelastic phase separation. We note
that we can observe the same regimes also in our numerical simulations
in \cite{Brunk.08.07.2019}, see Figure 1 of
Ref.~\cite{Brunk.08.07.2019} and the associated discussion.

\begin{figure*}
  \begin{minipage}{.5\linewidth}
    \centering
    \bfseries Mesoscopic
  \end{minipage}%
  \begin{minipage}{.5\linewidth}
    \centering
    \bfseries Macroscopic
  \end{minipage}

  \includegraphics{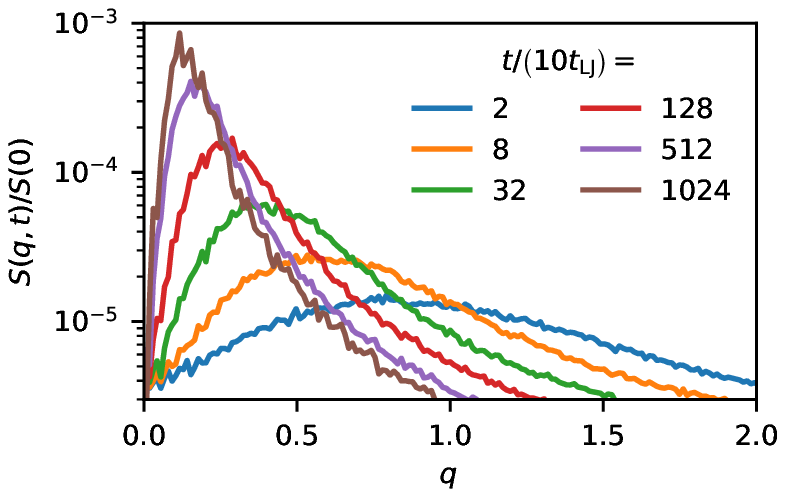}%
  \includegraphics{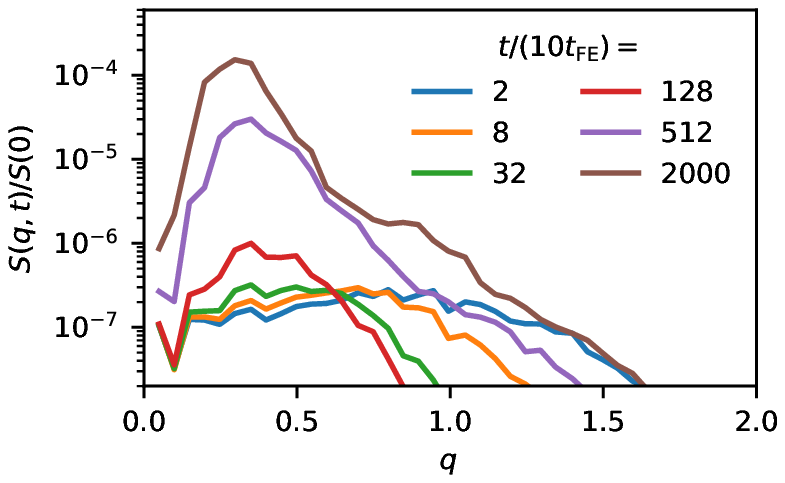}

  \includegraphics{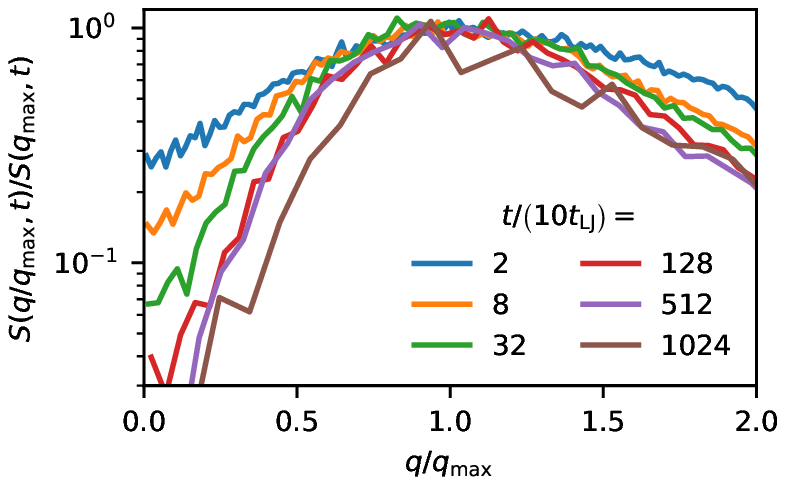}%
  \includegraphics{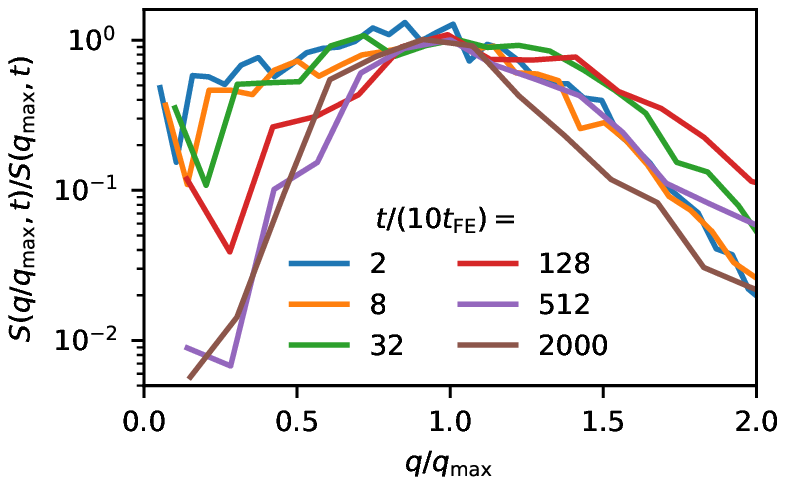}

  \includegraphics{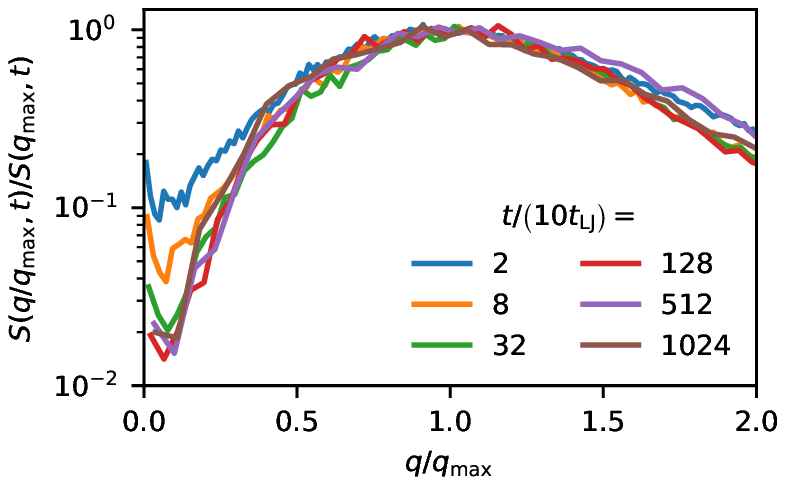}%
  \includegraphics{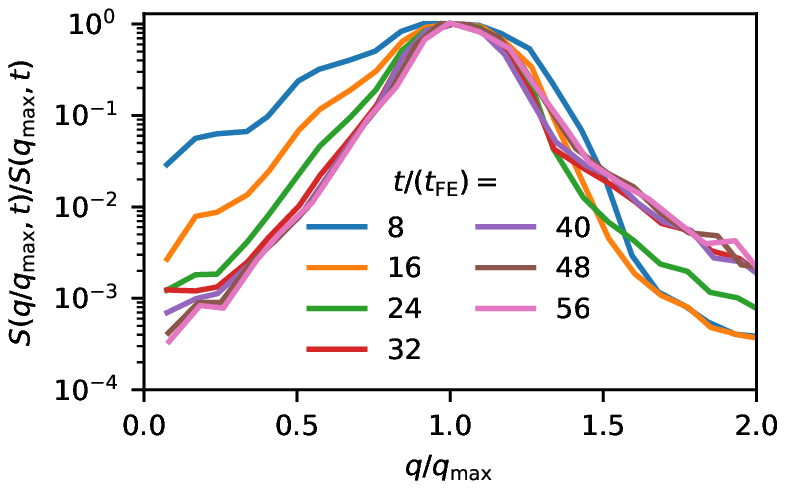}

  \caption{%
    Time evolution of the structure factor (left panel: mesoscopic
    simulations, right panel: macroscopic simulations).
    The first row shows the normalized scattering intensity $S(q,t)/S(q=0)$
    vs. the wavenumber $q$. Note that the normalization implies
      a value of unity at $q=0$. Sinced it would be completely
      off scale, this value is omitted.
    In the second row, the scattering intensity has been normalized by its peak
    value $S(q_\mrm{max})$ and the wavenumber by $q_\mrm{max}$.
    For comparison we show another normalized plot for
    simple fluids without any elastic effects in the third row.
  }\label{fig:struc}
\end{figure*}

Initial configurations were produced running the mesoscopic model at
temperature $T=1$ enforced by a Langevin thermostat with friction
constant $1$. For the initial equilibration, the polymer system was
not coupled to the LB fluid. The system consists of $1024$ polymer
chains, each of which comprises $128$ beads of mass $m=1$, in a box of
size $512\times 512\times 4$. At equilibration time the nonbonded
interaction is purely repulsive, corresponding to an attraction
strength of $\lambda=0$. The FENE bond potential has a strength of
$K=30$ and a maximum extension of $1.5$. In order to generate
configurations that are effectively two-dimensional, the particles are
confined in the $z$-direction by an external potential while making
sure that fluctuations in the $z$-direction are small compared to
$\sigma_\mrm{LJ}$. The LB fluid has shear and bulk viscosity of $3$
and its density is $1$; it is coupled to the particles with
a Stokes friction parameter of $20$. For every $10$ Molecular Dynamics
integrations, one LB step is performed. We extracted mean and variance
of the mesoscopic density and used this for a starting configuration
for the macroscopic simulations. In the mesoscopic model, phase
separation is induced by suddenly introducing an attractive well with
a depth of $\lambda=2$ in the non-bonded potential.

Then the structure factor as defined by
\begin{equation}
  S(\mathbf{q},t) = \snorm*{\int_\Omega e^{i\mathbf{q}\cdot x} \phi(x,t) \dx}^2
\end{equation}
was calculated for both approaches at successive times $t$.
More precisely, we average the structure factor over spherical shells, i.e.
\begin{equation}\label{eq:structure_factor_shell_avg}
  S(q;q+dq,t) = \int_{Z(q,dq)} S(\mathbf{q},t) d\mathbf{q}.
\end{equation}
Here $Z(q,dq):=\{q\in\mathbb{R}^d : q< \norm{\mathbf{q}}_2 \leq
q+dq\}$. In case of the mesoscopic model, configurations were
interpolated onto a discrete two-dimensional lattice in order to
evaluate the structure factor via Fast Fourier Transform. Four
independent simulations were performed and the average of the
structure factor curves taken. $25$ independent runs were used in case
of the macroscopic model.
Note that at $q = 0$ the structure factor is time-independent, because
of the conservation law $\int_\Omega \phi(x,t) \dx = \textrm{const.}$.

In Fig.~\ref{fig:struc} we can see the structure factors of both methods.
Larger systems allow for a higher resolution, as the smallest
resolvable wavenumber depends on system size $L$ via $q_0=2\pi/L$,
resulting in a better resolution for the mesoscopic simulation data.
One should note, however, that the mesoscopic simulations are orders of
magnitude more costly in terms of CPU time compared to the macroscopic
simulations.

The value of $\lim_{q\to 0}S(q)$ is very different from $S(q = 0)$; in
thermal equilibrium it is proportional to the compressibility. The
figures show that the compressibilities of the two models differ
substantially, because the equation of state was not yet adjusted to
match.
After the quench at $t=0$ a peak starts to develop, which
grows in height and moves towards smaller $q$ values as time
progresses.

In the second row of Fig.~\ref{fig:struc}, the scattering intensity is
normalized by its maximum value $S(q_\mrm{max})$ and the wavenumber
scaled by $q_\mrm{max}$. This fixes the peak at $(1,1)$ in the spirit
of the analysis of Tanaka and
Araki~\cite{tanaka_viscoelastic_2006}. The fact that the curves do
\emph{not} collapse onto a master curve indicates that the dynamic
scaling hypothesis is violated and the demixing behavior is
non-standard.  In order to confirm that finding, we have conducted
simulations of simple fluids and created an analogous plot.  In case
of the mesoscopic model, the same system was simulated without
connectivity, which is accommodated for by a larger quench depth of
$\lambda=3$.  For the macroscopic model elastic stresses where
neglected, i.e. the equation and the coupling terms for $q$ and $\CC$
are removed. Note that to neglect $q$ setting $A(\phi)=0$ is
sufficient. In row three of Fig.~\ref{fig:struc} we see that the
dynamic structure factor indeed shows a tendency to collapse for
simple fluids. Disagreement for small wave numbers is expected due to
finite-size effects.

Since the wavenumber with highest scattering intensity $q_{\max}$ is
inversely proportional to a characteristic length scale in the system,
$q_\mrm{max}(t)$ reflects the coarsening dynamics in phase separation.
\begin{figure}
  \centering
  \includegraphics{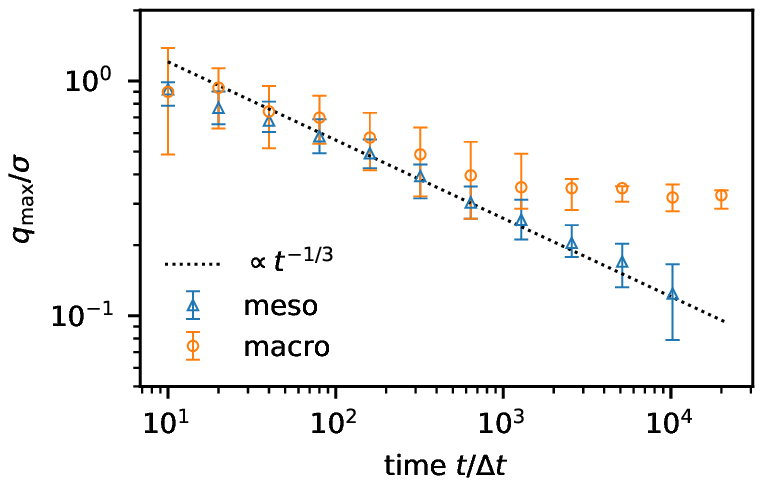}
  \caption{Time evolution of peak wavenumber
    $q_\mrm{max}$. Triangles indicate results from the mesoscopic
    model ($\Delta t = t_\mrm{LJ}, \sigma=\sigma_\mrm{LJ})$, while
    circles represent the macroscopic model ($\Delta t =t_{FE}$). For
    reference, the Lifshitz-Slyozov growth law $q_\mrm{max}\propto
    t^{-1/3}$ is given as well.  }
  \label{fig:peak_position}
\end{figure}
In Fig.~\ref{fig:peak_position} we show $q_\mrm{max}(t)$ as obtained
from mesoscopic and macroscopic simulations in a double logarithmic
plot. In the initial phase the agreement is quite good and both
approaches show a growth-law with an exponent close to the
Lifshitz-Slyozov exponent of ${-1/3}$, hinting at diffusive dynamics.
At about $t\approx 512 \Delta t$ however, the macroscopic data starts
to deviate and becomes flatter, indicating a transition to a different
dynamical regime.
\begin{figure}
  \centering
  \includegraphics{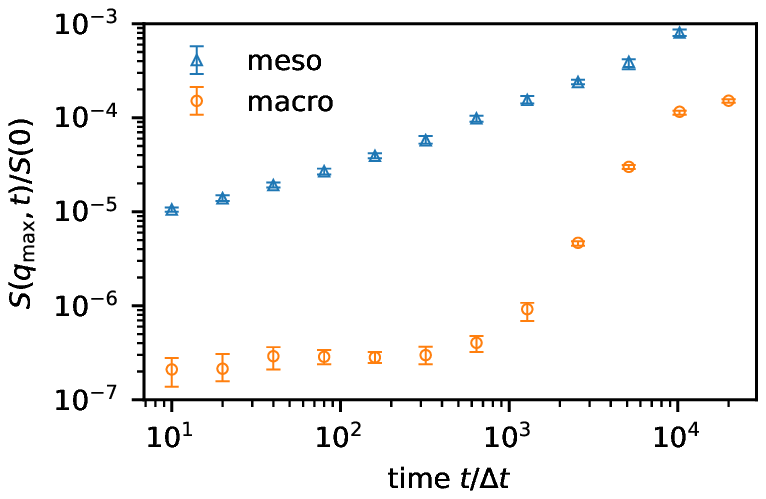}
  \caption{Normalized peak-scattering intensity $S(q_\mrm{max},t)/S(0,t)$ for
    mesoscopic ($\Delta t = t_\mrm{LJ}$) and macroscopic simulations
      ($\Delta t = t_\mrm{FE}$) respectively.
      The difference in scale is due to discrepancies in the equation of state
      of both approaches.
    }%
  \label{fig:peak_amplitude}
\end{figure}
This is confirmed by Fig.~\ref{fig:peak_amplitude} which shows a sharp rise in
scattering intensity at this point, whereas the behavior of the mesoscopic
simulation is more regular.

This deviation is not surprising, since the parameters of the
macroscopic model are not really calibrated to the mesoscopic
description. The peak wave number is highly determined by the
interplay of interface width $c_0$ and the choice of potential
$f(\phi)$. However $c_0$ is directly related to the length scale.  In
the future we want to calibrate the interface width and the potential
in order to obtain more accurate results. Furthermore, we want to
investigate the effects of the length scales by conducting a variety
of simulations based on different coarse-graining levels of the
mesoscopic data, i.e. using more data than only mean and variance.

\section{Discussion and outlook}

In this work we have first considered a suitable reduction of a
modified binary fluid model. The system is closed by deriving a
relation between the relative velocity and suitable state variables
that arise. In the course of this model reduction we obtain a relation
for the relative velocity that is quite similar to the one given in
\cite{Zhou.2006}. To arrive at this model, we make essentially the
same (probably partly debatable) phenomenological assumptions as
\cite{Zhou.2006}, but with a derivation that makes the elimination of
the relative velocity more explicit.  Furthermore, the construction
gives insight what the concrete physical interpretation of $q$
may be, since the interpretation of $\TT_v$ or $\CC$ is fairly
clear. Inclusion of dissipative terms and suitable evolution equations
for the viscoelastic effects lead to our new macroscopic model
\eqref{eq:full_model}.

In the second part we shortly discuss the thermodynamic consistency of
the proposed model and furthermore present our results on existence of
weak solution in two relevant cases. Afterwards we introduced the
notion of relative energy, as a suitable \emph{distance} in terms of
energy. Finally, we discussed the stability and weak-strong uniqueness
by means of relative energy methods. All together the section implies
 well-posedness of the problem.

In the last part we compared the results of the microscopic
description with our macroscopic model. We can see that the structure
factor is a valuable tool in comparing key features of both
approaches.

In order to perform a more thorough comparison however, the parameters
of both models need to be adjusted to match, and to do so the
interpretability of the macroscopic equations in terms of microscopic
physics needs to be improved. This was the main motivation for
parallel investigations by us, which aimed at the development of a
viscoelastic phase separation model from scratch, i.e. starting from a
simple microscopic model and then applying coarse-graining. This
alternative model, which has the advantage of being well-rooted in
microscopic physics, is presented in our companion paper
\cite{spiller2021systematic}, with analytical results that look quite
encouraging. While the equations share some similarities with those
presented here, we note that the bulk stress is not present and the
energetic structure is different. Whether the closure of the present
paper can be justified from microscopic physics remains an open
question, into which we will look further in the future. We also plan
to study the differences between the models in order to deepen our
understanding of the problem.

\ack

Funded by the Deutsche Forschungsgemeinschaft (DFG, German Research
Foundation), Project No. 233630050-TRR 146.

\onecolumn

\section*{References}

\begin{flushleft}
\bibliographystyle{iopart-num}
\bibliography{bibstuff}
\end{flushleft}

\end{document}